\documentclass{emulateapj}
\usepackage{float}
\usepackage{amsmath}

\def\app#1#2{%
  \mathrel{%
    \setbox0=\hbox{$#1\sim$}%
    \setbox2=\hbox{%
      \rlap{\hbox{$#1\propto$}}%
      \lower1.1\ht0\box0%
    }%
    \raise0.25\ht2\box2%
  }%
}

\begin{document}

\title{Stellar Velocity Dispersion: Linking Quiescent Galaxies to their Dark Matter Halos }

\author{H. Jabran Zahid, Jubee Sohn \& Margaret J. Geller} 
\affil{Smithsonian Astrophysical Observatory, Harvard-Smithsonian Center for Astrophysics - 60 Garden Street, Cambridge, MA 02138}

\def\mean#1{\left< #1 \right>}

\begin{abstract}

We analyze the Illustris-1 hydrodynamical cosmological simulation to explore the stellar velocity dispersion of quiescent galaxies as an observational probe of dark matter halo velocity dispersion and mass. Stellar velocity dispersion is proportional to dark matter halo velocity dispersion for both central and satellite galaxies. The dark matter halos of central galaxies are in virial equilibrium and thus the stellar velocity dispersion is also proportional to dark matter halo mass. This proportionality holds even when a line-of-sight aperture dispersion is calculated in analogy to observations. In contrast, at a given stellar velocity dispersion, the dark matter halo mass of satellite galaxies is smaller than virial equilibrium expectations. This deviation from virial equilibrium probably results from tidal stripping of the outer dark matter halo. Stellar velocity dispersion appears insensitive to tidal effects and thus reflects the correlation between stellar velocity dispersion and dark matter halo mass prior to infall. {There is a tight relation ($\lesssim0.2$ dex scatter) between line-of-sight aperture stellar velocity dispersion and dark matter halo mass suggesting that the dark matter halo mass may be estimated from the measured stellar velocity dispersion for both central and satellite galaxies. We evaluate the impact of treating all objects as central galaxies if the relation we derive is applied to a statistical ensemble. A large fraction ($\gtrsim 2/3$) of massive quiescent galaxies are central galaxies and systematic uncertainty in the inferred dark matter halo mass is $\lesssim 0.1$ dex thus simplifying application of the simulation results to currently available observations.}

\end{abstract}

\section{Introduction}

\begin{deluxetable*}{cccccc}
\tablecaption{$\mathrm{log}(y) = \alpha + \beta \mathrm{log}(x)$}
\tablehead{\colhead{$y$} & \colhead{$x$} & \colhead{$\alpha$} & \colhead{$\beta$} & \colhead{RMS (dex)} & \colhead{Figure}}
\startdata
$\frac{\sigma_{T, DM}}{100 ~\mathrm{km ~ s}^{-1}}$ & $\frac{M_{DM}}{10^{12} M_\odot}$ & \!\!\!\!$-0.007 \pm 0.001$  & $0.300 \pm 0.001$ & 0.02 &1A\\ \\
$\frac{\sigma_{T,DM}}{100 ~\mathrm{km ~ s}^{-1}}$ & $\frac{\sigma_{T, \ast}}{100 ~\mathrm{km ~ s}^{-1}}$ & $0.023 \pm 0.004$ & $1.06 \pm 0.02$ & 0.14 & 1B  \\ \\
$\frac{M_{DM}}{10^{12} M_\odot}$              & $\frac{\sigma_{T, \ast}}{100 ~\mathrm{km ~ s}^{-1}}$ & $0.09 \pm 0.02$ & $3.48 \pm 0.05$ & 0.13 & 2  \\ \\
$\frac{\sigma_{T,DM}}{100 ~\mathrm{km ~ s}^{-1}}$ & $\frac{\sigma_{T, \ast}}{100 ~\mathrm{km ~ s}^{-1}}$ & $0.041 \pm 0.003$ & $1.00 \pm 0.01$ & 0.14 & 3  \\ \\
$\frac{\sigma_{h, \ast}}{100 ~\mathrm{km ~ s}^{-1}}$ & $\frac{\sigma_{T, \ast}}{100 ~\mathrm{km ~ s}^{-1}}$ & $0.01\pm0.01$ & $0.96\pm0.02$ & 0.04 & 5A  \\ \\
$\frac{M_{DM}}{10^{12} M_\odot}$              & $\frac{\sigma_{h, \ast}}{100 ~\mathrm{km ~ s}^{-1}}$ & $0.16\pm0.03$& $3.31\pm0.10$ & 0.17 & 5B \\ 
\enddata
\tablecomments{Parameters for various linear fits presented in this work. Columns 1 and 2 list the dependent and independent variable, respectively. Columns 3 and 4 give the zero-point and power law slope of each fit, respectively. Scatter around the best-fit relation is given in column 5 and column 6 provides the figure reference for each fit.}
\label{tab:stack_prop}
\end{deluxetable*}

Dark matter is an enduring mystery. Despite its ubiquity, dark matter is not directly detected and its properties are largely constrained from observations of baryons. A galaxy composed of baryons forms at the center of a dark matter halo and the two coevolve. Thus, galaxies are directly observed objects which trace the evolving dark matter distribution in the Universe. Linking galaxies and their enigmatic dark matter halos is critical for understanding structure formation. 

Various techniques have been developed to link galaxies to their dark matter halos \citep[and references therein]{Moster2010}. A standard approach takes measurable galaxy properties (e.g., luminosity, stellar mass, velocity dispersion) and links them to dark matter halos assuming a one-to-one rank order correspondence with halo mass \citep[i.e. abundance matching;][]{Yang2003, Kravtsov2004, Tasitsiomi2004, Conroy2006, Berrier2006, Shankar2006, Guo2010, Behroozi2013a}. Stellar mass is commonly provides a quantitative connection between galaxies and dark matter halos over cosmic time \citep{Moster2010, Behroozi2010}. Mass loss of subhalos due to tidal stripping obfuscates the connection and must be treated \citep{Vale2006, Conroy2006}. The abundance matching approach based on stellar mass is subject to both theoretical and observational uncertainties and limitations \citep[e.g.,][]{Behroozi2010, Campbell2017}.

Central stellar velocity dispersion is an observable property of galaxies which reflects the gravitational potential. Several studies suggest that velocity dispersion is a fundamental observable characterizing galaxies and their dark matter halos \citep{Wake2012b, Bogdan2015, Schechter2015, Zahid2016c}. Unlike stellar mass, stellar velocity dispersion is a directly measured quantity related to the gravitational potential; systematic uncertainties in stellar velocity dispersion of quiescent galaxies are $<0.03$ dex \citep{Fabricant2013, Zahid2016a, Zahid2016c}. The small systematic uncertainties and the straightforward physical interpretation make velocity dispersion an attractive alternative to stellar mass for connecting galaxies to their dark matter halos. 

{\citet{Schechter2015} argues that the central stellar velocity dispersion of a galaxy is a good proxy for the velocity dispersion of its dark matter halo.} \citet{Zahid2016c} show that the relation between stellar velocity dispersion and total mass (dark + baryonic) for galaxies and massive galaxy clusters is consistent with the theoretical relation for dark matter halos \citep[see][]{Evrard2008, Rines2016}. {\citet{Schechter2015} and \citet{Zahid2016c} suggest a nearly one-to-one correspondence between the central stellar velocity dispersion and the velocity dispersion of the dark matter halo. Thus, stellar velocity dispersion could be an observable property directly related to the dark matter halo.} 

Several studies have examined whether the velocity dispersion of galaxies in clusters is an unbiased proxy for the velocity dispersion of the cluster dark matter halo \citep[e.g.,][]{Biviano2006, Faltenbacher2006, Lau2010, Munari2013, Rines2016, Ntampaka2017, Armitage2018}. In analogy to these studies, here we examine results from a hydrodynamical cosmological simulation to test theoretically whether the central stellar velocity dispersion of a galaxy is a good proxy for the velocity dispersion of its dark matter halo.

We explore the connection between the stellar velocity dispersion of quiescent galaxies and their dark matter halo velocity dispersion and mass using the Illustris hydrodynamical cosmological simulation \citep{Springel2010, Vogelsberger2014a}. The Illustris galaxy formation model is not explicitly tuned to reproduce the kinematic properties of galaxies \citep{Vogelsberger2013}. The free parameters of the physical model are tuned to reproduce the observed stellar-to-halo mass relation \citep{Vogelsberger2013}. In contrast to stellar mass, stellar velocity dispersion is likely to be insensitive to the baryonic physical model. Thus, the simulation is well-suited for exploring the connection between stellar velocity dispersion and dark matter halo properties. We describe the simulations and methods in Section 2 and present results in Section 3. We discuss our results and their connection to the observations in Section 4. We conclude in Section 5.

\begin{figure*}
\begin{center}
\includegraphics[width = 2 \columnwidth]{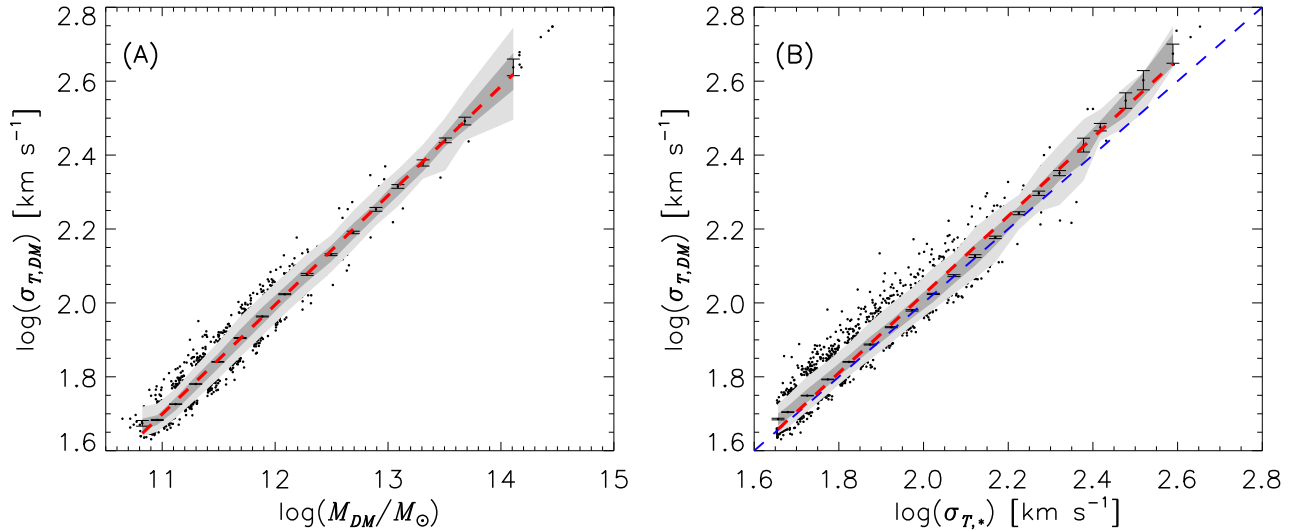}
\end{center}
\caption{(A) Total dark matter halo velocity dispersion, $\sigma_{T,DM}$, as a function of dark matter halo mass, $M_{DM}$, for central galaxies. Dark and light gray bands denote the central 50 and 90\% of the $\sigma_{T,DM}$ distribution, respectively.  The red dashed line is the best-fit relation to the median $\sigma_{T,DM}$ calculated in bins of $M_{DM}$. Error bars are bootstrapped. (B) $\sigma_{T,DM}$ as a function of total stellar velocity dispersion, $\sigma_{T,\ast}$ for central galaxies. Dark and light gray bands denote the central 50 and 90\% of the $\sigma_{T,DM}$ distribution, respectively. The red dashed line is the best-fit relation to the median $\sigma_{T,DM}$ calculated in bins of $\sigma_{T,\ast}$. Error bars are bootstrapped. The blue dashed line is one-to-one correspondence.}
\label{fig:virial}
\end{figure*}

\section{Simulation Data and Methods}

\subsection{Illustris-1 Simulation}

The Illustris Project is a set of large-scale hydrodynamical cosmological simulations of galaxy formation and evolution \citep{Vogelsberger2014a} using the moving mesh code AREPO \citep{Springel2010}. We analyze the highest resolution Illustris-1 simulation which is publicly available\footnote{http://www.illustris-project.org/} \citep{Nelson2015}. The simulation volume is $1.2 \times 10^{6}$ Mpc$^{3}$ with a dark matter and initial baryonic mass resolution of $6.3 \times 10^6$ $M_\odot$ and $1.3 \times 10^6$ $M_\odot$, respectively. The simulation is run with a full galaxy formation model \citep{Vogelsberger2013} and the \emph{Wilkinson Microwave Anisotropy Probe} 9 cosmology \citep{Hinshaw2013}: $\Omega_m = 0.2726$, $\Omega_\Lambda = 0.7274$, $\Omega_b = 0.0456$, $H_0 = 100h~ \mathrm{km ~ s}^{-1} ~ \mathrm{Mpc}^{-1}$. We convert all relevant quantities to $h=0.704$.

Simulated particle data is publicly available in 136 snapshots between $0<z<47$. For each snapshot, a friends-of-friends algorithm is used to identify dark matter halos \citep{Davis1985} and the SUBFIND algorithm is used to identify gravitationally bound sub-structures which are associated with individual galaxies \citep{Springel2001}. We analyze the $z=0$ galaxy population but trace individual galaxies residing in larger substructures back to the snapshot corresponding to redshift just prior to infall. We refer to the corresponding infall redshift as $z_{in}$. We calculate $z_{in}$ of a z=0 satellite galaxy as the snapshot where it was last a central galaxy. For satellite galaxies in our sample, the infall redshift is $0.01 < z_{in} <2.7$.

{The dark matter halo mass, $M_{DM}$, corresponds to the mass of a single gravitationally bound structure and does not include mass in bound substructures. In practice, this distinction makes little difference as the mass in substructures is typically a negligible fraction of $M_{DM}$. For satellite galaxies, $M_{DM}$ can be biased low if the subhalo is close to peri-center on its orbit (V. Springel, private communication). This bias probably impacts only a small fraction of the sample.}

Our aim is to analyze simulations in a manner consistent with observations. Thus, we select quiescent galaxies at $z=0$ by requiring a specific star formation rate $<2\times10^{-10}$ yr$^{-1}$ \citep{Wellons2015a}. {This selection is analogous to an observational sample selection which excludes late-type galaxies with stellar systems dominated by ordered rotation rather than random motions.} To avoid resolution limitations and to minimize selection bias we analyze galaxies with dark matter halo masses $>10^{10.4} M_\odot$ and total and aperture stellar velocity dispersions $>45$ km s$^{-1}$. We cut in total and aperture stellar velocity dispersions because both these quantities are taken as independent variables in this work. These selection criteria yield a sample of 12467 subhalos hosting quiescent galaxies at $z=0$; 8998 of these subhalos are central galaxies. The minimum number of stellar and dark matter particles for our selected sample is 1234 and 4053, respectively. {Even for the least massive halos, the number of stellar and dark matter particles is well above the resolution limit required to derive a robust velocity dispersion (e.g., \citealt{Evrard2008} suggest at least 45 particles are needed).}

\subsection{Velocity Dispersion}

We calculate total velocity dispersion of all dark matter and star particles associated with each subhalo by summing the velocity dispersion along the three principle axes in quadrature:
\begin{equation}
\sigma_{T} =  \sqrt { \frac{\sigma_x^2 + \sigma_y^2 + \sigma_z^2}{3} }.
\end{equation}
Here velocity dispersion is the square root of the variance of velocities using the standard mathematical definition. Dark matter particles are equal mass but stellar particles are not. To account for unequal stellar particle contributions and for consistency with observations, we appropriately weight each stellar particle by its luminosity in the $g-$band when calculating velocity dispersion. Stellar velocity dispersion is typically measured in the wavelength range $4000 \sim 6000\mathrm{\AA}$ \citep{Fabricant2013, Thomas2013} which roughly corresponds to the $g-$band.

The total three dimensional stellar velocity dispersion can not be measured observationally. Instead, central stellar velocity dispersion measured using fiber spectroscopy is a line-of-sight measurement of all stars within a cylinder. We calculate a consistent measure of stellar velocity dispersion using the simulation data. We take the center of the stellar mass distribution within each subhalo, $[x_c, y_c, z_c]$, and calculate the two dimensional projected radius on the sky for the $i$th particle as
\begin{equation}
R_{xy, i} = \sqrt{  (x_i - x_c)^2 + (y_i - y_c)^2 }.
\end{equation}
Here $x,y,z$ are the cartesian physical coordinate positions of each particle. We calculate the line-of-sight velocity dispersion, $\sigma_z$, for all particles within the circular aperture $R_{xy}< R_h$. Here $R_h$ is the two-dimensional projected stellar half-light radius calculated in analogy to observations. We refer to this line-of-sight aperture velocity dispersion measurement as $\sigma_{h}$.

\section{Relation Between Stellar Velocity Dispersion and Dark Matter Halo Properties}

\begin{figure}
\begin{center}
\includegraphics[width =  \columnwidth]{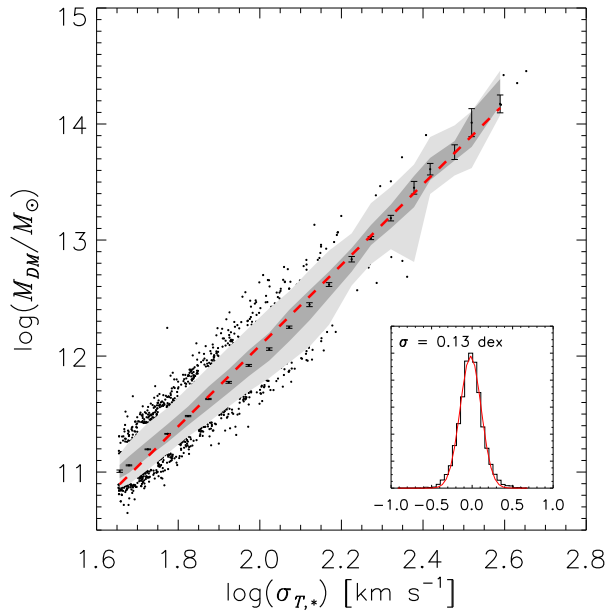}
\end{center}
\caption{Dark matter halo mass, $M_{DM}$, as a function of total stellar velocity dispersion, $\sigma_{T,\ast}$, for central galaxies. Dark and light gray bands denote the central 50 and 90\% of the $M_{DM}$ distribution, respectively. The red dashed line is the best-fit relation to the median $M_{DM}$ calculated in bins of $\sigma_{T,DM}$. Error bars are bootstrapped. The inset shows a histogram of the residuals with a Gaussian over-plotted in red. }
\label{fig:vm}
\end{figure}

\begin{figure*}
\begin{center}
\includegraphics[width = 1.8 \columnwidth]{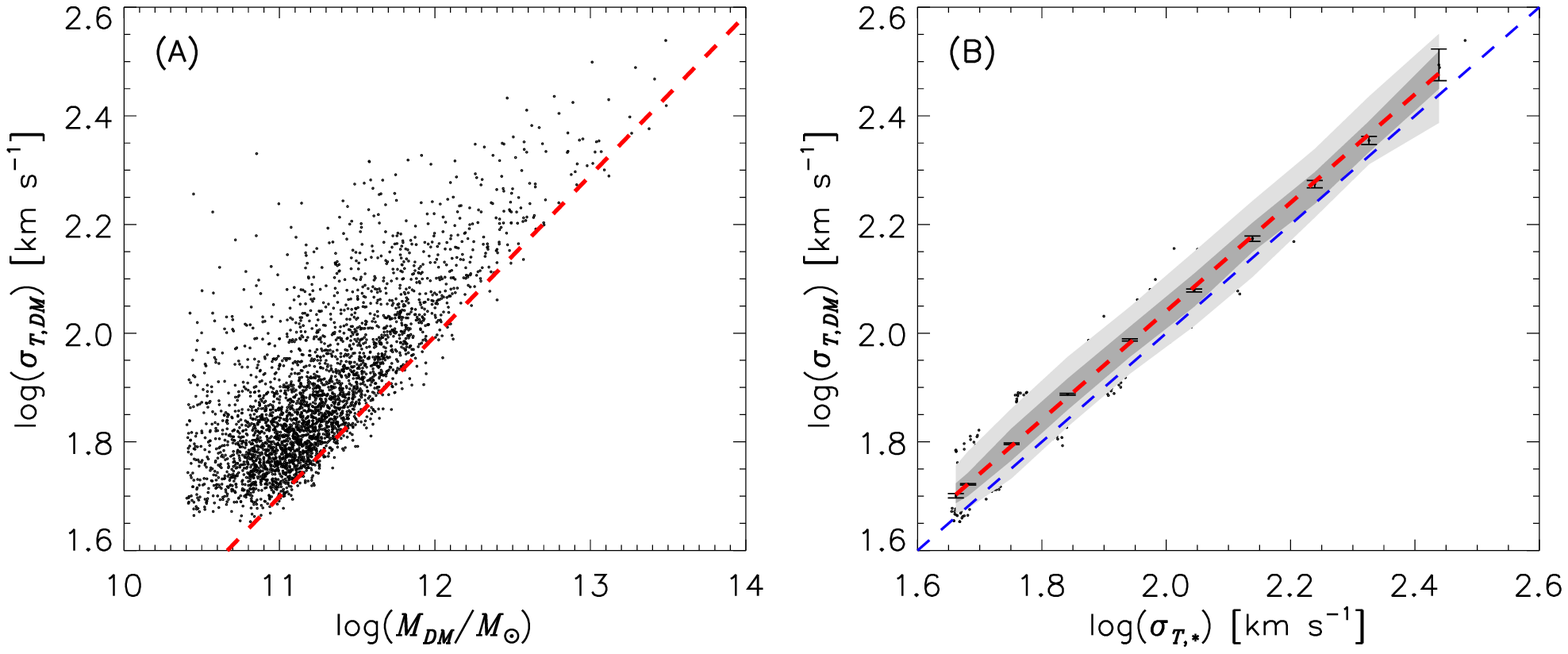}
\end{center}
\caption{A) Total dark matter halo velocity dispersion, $\sigma_{T,DM}$, as a function of dark matter halo mass, $M_{DM}$, for satellite galaxies. The red dashed line is the best-fit relation for central galaxies shown in Figure \ref{fig:virial}A. (B) $\sigma_{T,DM}$ as a function of total stellar velocity dispersion, $\sigma_{T,\ast}$ for satellite galaxies. Dark and light gray bands denote the central 50 and 90\% of the $\sigma_{T,DM}$ distribution, respectively. The red dashed line is the best-fit relation to the median $\sigma_{T,DM}$ calculated in bins of $\sigma_{T,\ast}$. Error bars are bootstrapped. The blue dashed line is one-to-one correspondence.}
\label{fig:vsat}
\end{figure*}

To establish the physical basis of the connection between stellar velocity dispersion and dark matter halo velocity dispersion and mass we analyze total velocity dispersions which are theoretical quantities. We first examine central galaxies and then satellite galaxies. To connect theoretical results to observations, we derive the relations as a function of line-of-sight aperture stellar velocity dispersion. We show that indeed the derived relations are robust to the way stellar velocity dispersion is calculated.

\subsection{Theoretical Relations for Central Galaxies}

Figure \ref{fig:virial}A shows the relation between total dark matter halo velocity dispersion, $\sigma_{T,DM}$, and dark matter halo mass, $M_{DM}$. We fit the binned relation using \emph{linfit.pro} in IDL: 
\begin{equation}
\mathrm{log}\left(  \frac{\sigma_{T,DM}}{100 ~\mathrm{km ~ s}^{-1}}   \right)  = \alpha_1 + \beta_1 \mathrm{log}\left( \frac{M_{DM}}{10^{12} M_\odot} \right).
\end{equation}
The best-fit parameters are $\alpha_1 =  -0.007 \pm 0.001$ and $\beta_1 = 0.300 \pm 0.001$; the fit errors quoted throughout reflect the dispersion in the binned data. The relation has a root-mean-square (RMS) scatter of $\sim0.02$ dex. The tight relation and scaling of the two parameters is a consequence of virial equilibrium \citep[c.f.,][]{Evrard2008}. 

Figure \ref{fig:virial}B shows the relation between $\sigma_{T,\ast}$ and $\sigma_{T,DM}$. The best-fit relation is
\begin{equation}
\mathrm{log}\left(  \frac{\sigma_{T,DM}}{100 ~\mathrm{km ~ s}^{-1}}   \right)= \alpha_2 + \beta_2\mathrm{log}\left(  \frac{\sigma_{T,\ast}}{100 ~\mathrm{km ~ s}^{-1}}   \right)
\end{equation}
with $\alpha_2 = 0.023 \pm 0.004$, $\beta_2 = 1.06 \pm 0.02$ and an RMS scatter of 0.14 dex. On average, $\sigma_{T,DM}$ is nearly directly proportional to $\sigma_{T,\ast}$. Thus, the stellar velocity dispersion is a robust tracer of the dark matter halo velocity dispersion.

The consistency between $\sigma_{T,\ast}$ and $\sigma_{T,DM}$ implies that for central galaxies, $\sigma_{T,\ast}$ is a tracer of $M_{DM}$. Figure \ref{fig:vm} shows the relation between $\sigma_{T,\ast}$ and $M_{DM}$. The best-fit relation is 
\begin{equation}
\mathrm{log}\left(  \frac{M_{DM}}{10^{12} ~ M_\odot}   \right) = \alpha_3+ \beta_3 ~ \mathrm{log}\left(  \frac{\sigma_{T,\ast}}{100 ~\mathrm{km ~ s}^{-1}}   \right) 
\label{eq:sdm}
\end{equation}
with $\alpha_3 =  0.09 \pm 0.02$, $\beta_3 = 3.48 \pm 0.05$ and RMS scatter of 0.13 dex. The scaling between  $\sigma_{T,\ast}$ and $M_{DM}$ is consistent with virial equilibrium. This consistency is a consequence of the nearly direct proportionality between $\sigma_{T,\ast}$ and $\sigma_{T,DM}$ (see Figure \ref{fig:virial}B).

\subsection{Theoretical Relations for Satellite Galaxies}

For central galaxies $\sigma_{T,\ast}$ is a proxy of $M_{DM}$. Here we derive theoretical relations for the satellite galaxy population.

{Figure \ref{fig:vsat}A shows the relation between $\sigma_{T,DM}$ and $M_{DM}$ for satellite galaxies. The red dashed line is the virial equilibrium relation for central galaxies (Figure \ref{fig:virial}A). Dark matter halos of satellite galaxies depart from virial equilibrium. }

{Figure \ref{fig:vsat}B shows the relation between $\sigma_{T,\ast}$ and $\sigma_{T,DM}$ for these satellite galaxies. The best-fit relation is
\begin{equation}
\mathrm{log}\left(  \frac{\sigma_{T,DM}}{100 ~\mathrm{km ~ s}^{-1}}   \right)= \alpha_4 + \beta_4 \mathrm{log}\left(  \frac{\sigma_{T,\ast}}{100 ~\mathrm{km ~ s}^{-1}}   \right)
\end{equation}
with $\alpha_4 = 0.041 \pm 0.003$, $\beta_4 = 1.00 \pm 0.01$ and an RMS scatter of 0.14 dex. On average, $\sigma_{T,DM}$ is directly proportional to $\sigma_{T,\ast}$. Thus, $\sigma_{T,\ast}$  is a robust tracer of $\sigma_{T,DM}$ for satellite galaxies despite the departures from virial equilibrium shown in Figure \ref{fig:vsat}A. The best-fit parameters of the relation between $\sigma_{T,\ast}$ and $\sigma_{T,DM}$ for satellite galaxies differ ($\sim 3 \sigma$) from the relation for central galaxies. }

\begin{figure*}
\begin{center}
\includegraphics[width =  1.65\columnwidth]{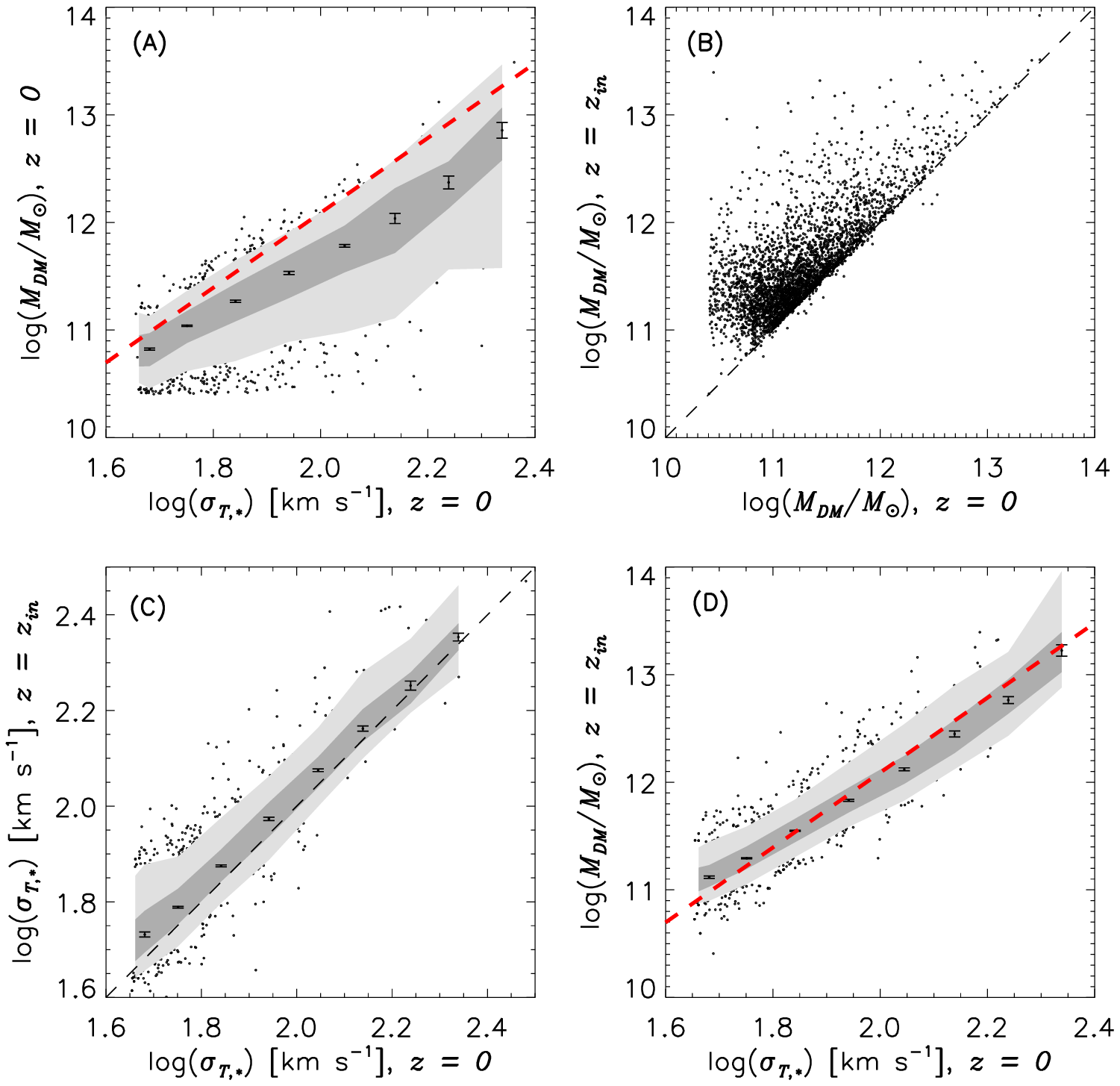}
\end{center}
\caption{(A) Dark matter halo mass, $M_{DM}$, as a function of total stellar velocity dispersion, $\sigma_{T,\ast}$, for satellite galaxies. Dark and light gray bands denote the central 50 and 90\% of the $M_{DM}$ distribution, respectively.  The red dashed line is the best-fit relation for central galaxies (Equation \ref{eq:sdm} and Figure \ref{fig:vm}). Points with error bars are the median $M_{DM}$ calculated in bins of $\sigma_{T,\ast}$ and errors are bootstrapped. (B) $M_{DM}$ at time of satellite infall, $z_{in}$, plotted against $M_{DM}$ at $z=0$ for satellite galaxies. The dashed line is one-to-one correspondence. (C) $\sigma_{T,\ast}$ at time of infall plotted against $\sigma_{T,\ast}$ at $z=0$ for satellite galaxies. Dark and light gray bands denote the central 50 and 90\% of the $\sigma_{T,\ast}$ at time of infall distribution, respectively. Points with error bars are the median $\sigma_{T,\ast}$ calculated in bins of $\sigma_{T,\ast}$ and errors are bootstrapped. The dashed line is one-to-one correspondence. (D) $M_{DM}$ at time of infall as a function of $\sigma_{T,\ast}$ at $z=0$ for satellite galaxies. Dark and light gray bands denote the central 50 and 90\% of the $\sigma_{T,\ast}$ at time of infall distribution, respectively. The red dashed line is the best-fit relation for central galaxies (Equation \ref{eq:sdm} and Figure \ref{fig:vm}). Points with error bars are the median $M_{DM}$ at time of infall calculated in bins of $\sigma_{T,\ast}$ and errors are bootstrapped. Note the consistency between the relation for satellite galaxies and the best-fit relation for central galaxies.}
\label{fig:sat}
\end{figure*}

\begin{figure*}
\begin{center}
\includegraphics[width =  1.83\columnwidth]{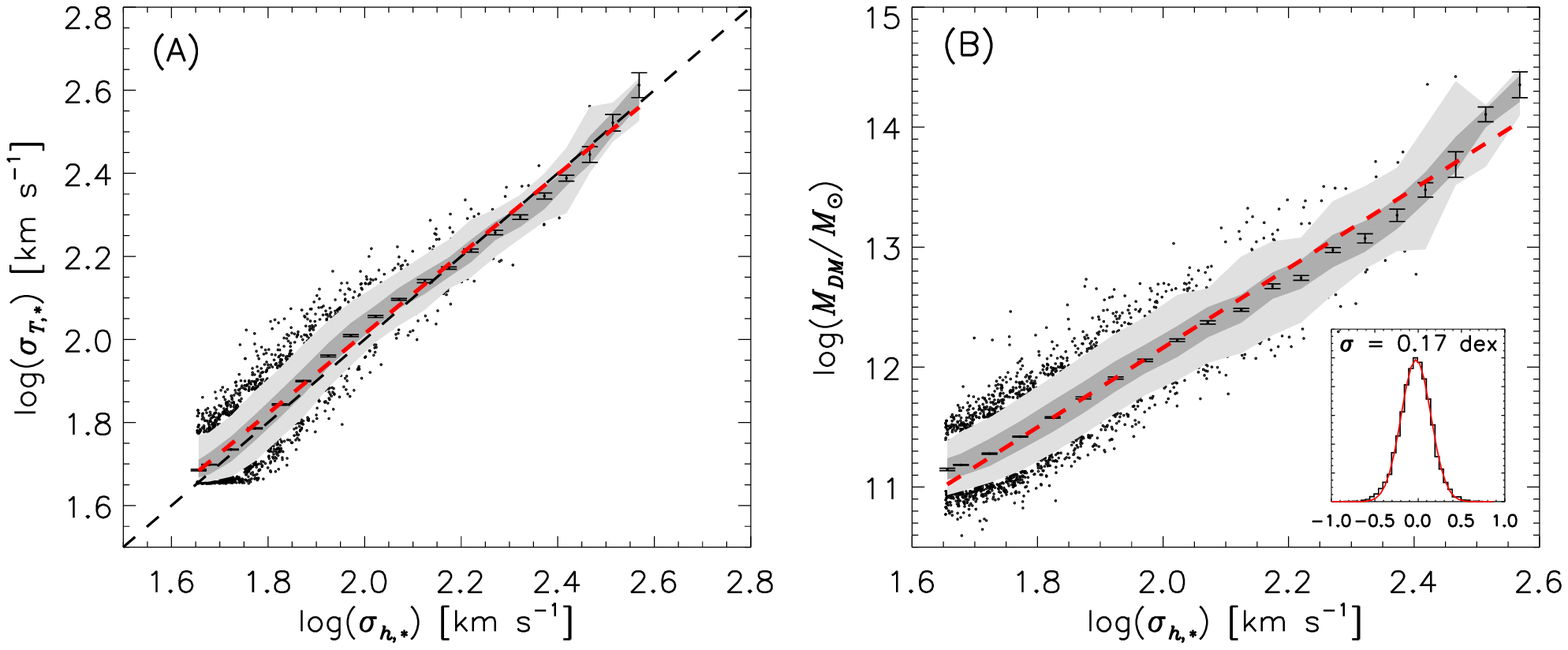}
\end{center}
\caption{(A) Line-of-sight stellar velocity dispersion measured within the half-light radius, $\sigma_{h,\ast}$, as a function of total stellar velocity dispersion, $\sigma_{T,\ast}$, for both central and satellite galaxies. Dark and light gray bands denote the central 50 and 90\% of the $\sigma_{h,\ast}$ distribution, respectively. The red dashed line is the best-fit relation to the median $\sigma_{T,\ast}$ calculated in bins of $\sigma_{h,\ast}$. Error bars are bootstrapped. The dashed line is one-to-one correspondence. (B) Dark matter halo mass, $M_{DM}$, as a function of $\sigma_{h,\ast}$ for both central and satellite galaxies. For satellite galaxies, $M_{DM}$ is the halo mass at time of infall. Dark and light gray bands denote the central 50 and 90\% of the $M_{DM}$ distribution, respectively. The red dashed line is the best-fit relation to the median $M_{DM}$ calculated in bins of $\sigma_{h,\ast}$. Error bars are bootstrapped. The inset shows a histogram of the residuals with a Gaussian plotted in red. }
\label{fig:v3}
\end{figure*}

Figure \ref{fig:sat}A shows the relation between $\sigma_{T,\ast}$ and $M_{DM}$ for satellite galaxies. The red dashed line is the relation for central galaxies (Figure \ref{fig:vm} and Equation \ref{eq:sdm}). Satellite galaxies are offset from the relation derived for central galaxies; at a given $\sigma_{T,\ast}$, $M_{DM}$ is smaller for satellite galaxies.

The offset in $M_{DM}$ of the satellite galaxy population likely results from tidal stripping. To test this scenario, we examine properties of each satellite galaxy at the redshift just before infall, $z_{in}$. Figure \ref{fig:sat}B shows $M_{DM}$ of each satellite at $z_{in}$ compared with $M_{DM}$ at $z=0$. The dark matter halo mass is more massive at the time of infall, supporting the scenario that dark matter halos of satellite galaxies are stripped.

Figure \ref{fig:sat}C shows the relation between $\sigma_{T,\ast}$ at $z_{in}$ and $\sigma_{T,\ast}$ at $z=0$. At $\sigma_{T,\ast} \lesssim 100$ km s$^{-1}$, the stellar velocity dispersion at $z_{in}$ is larger by $\lesssim0.05$ dex; for galaxies with $\sigma_{T,\ast} \gtrsim 100$ km s$^{-1}$, the two relations are consistent. Stellar velocity distributions of less massive galaxies may be more susceptible to external disturbance due to their shallower potential wells. This scenario is consistent with Figure \ref{fig:sat}B which shows more significant changes in dark matter halo masses for low mass galaxies. Differences in velocity dispersion are small even for galaxies at low velocity dispersion; {these differences result in the $\sim 3 \sigma$ difference between relations in Figure \ref{fig:virial}B and Figure \ref{fig:vsat}B. We conclude that the stellar and dark matter halo velocity dispersion, in contrast to the halo mass, is insensitive to tidal effects.}

The $\sigma_{T,\ast}$ of central galaxies is a proxy of $M_{DM}$. Figure \ref{fig:sat}D shows the relation between $\sigma_{T,\ast}$ at $z=0$ and $M_{DM}$ at $z_{in}$ for satellite galaxies. The red dashed line is the relation between $\sigma_{T,\ast}$ and $M_{DM}$ for central galaxies (note this line is not a fit to the data). For satellite galaxies, the relation between $\sigma_{T,\ast}$ and $M_{DM}$ at the time of infall is consistent with the relation for central galaxies. Thus, for satellite galaxies, stellar velocity dispersion reflects properties of the dark matter halo at the time of infall.

\subsection{Connecting Theory to Observations}

Here we analyze stellar velocity dispersion as an observed rather than as a theoretical quantity. Figure \ref{fig:v3}A shows $\sigma_{T,\ast}$ as a function of line-of-sight aperture stellar velocity dispersion measured within the half-light radius, $\sigma_{h,\ast}$. $\sigma_{h,\ast}$ is measured in a manner consistent with observations. The relation between $\sigma_{T,\ast}$ and $\sigma_{h,\ast}$ is
\begin{equation}
\mathrm{log}\left(  \frac{\sigma_{T,\ast}}{100 ~\mathrm{km ~ s}^{-1}}   \right) = \alpha_5 + \beta_5 \mathrm{log}\left(  \frac{\sigma_{h,\ast}}{100 ~\mathrm{km ~ s}^{-1}}   \right)
\end{equation}
with $\alpha_5 =  0.01\pm 0.01$, $\beta_5 = 0.96 \pm 0.02$ and RMS scatter of 0.04 dex; $\sigma_{h,\ast}$ is a robust proxy of $\sigma_{T,\ast}$ over most of stellar velocity dispersion range we probe. 

The consistency between $\sigma_{h,\ast}$ and $\sigma_{T,\ast}$ means that the observed velocity dispersion can be used to connect galaxies to their dark matter halos. Figure \ref{fig:v3}B shows the relation between $\sigma_{h,\ast}$ and $M_{DM}$. The best-fit is
\begin{equation}
\mathrm{log}\left(  \frac{M_{DM}}{10^{12} ~ M_\odot}   \right) = \alpha_6+ \beta_6 ~ \mathrm{log}\left(  \frac{\sigma_{h,\ast}}{100 ~\mathrm{km ~ s}^{-1}}   \right) 
\label{eq:sigma_dm}
\end{equation}
with $\alpha_6 =  0.16\pm 0.03$, $\beta_6 = 3.31 \pm 0.10$ and RMS scatter of 0.17 dex. This relation links the stellar velocity dispersion---a direct observable---to the dark matter halo mass.

For galaxies with large velocity dispersion, $\sigma_{h,\ast}$ systematically deviate from $\sigma_{T,\ast}$. The tangential component of the velocity distribution is greater for galaxies with large velocity dispersions. Aperture line-of-sight velocity dispersion is mostly sensitive to the radial velocity component and is biased towards smaller values as compared to the total velocity dispersion when the tangential velocity component is significant. The deviation between $\sigma_{h,\ast}$ and $\sigma_{T,\ast}$ seen in Figure \ref{fig:v3}A results from velocity anisotropy. The relation between $\sigma_{T,\ast}$---which is insensitive to anisotropy---and $M_{DM}$ does not deviate from a single power law fit at large velocity dispersions. Thus, the line-of-sight stellar velocity dispersion traces dark matter halo mass but velocity anisotropy is a source of systematic uncertainty and bias. This uncertainty affects all kinematic mass estimators based on line-of-sight velocity measurements \citep[e.g.,][]{Courteau2014}.

\subsection{Estimating Halo Mass from Velocity Dispersion}

We examine stellar velocity dispersion for central and satellite galaxies independently and show that stellar velocity dispersion is insensitive to stripping; the dark matter halo can however be stripped in dense environments. Thus, stellar velocity dispersion of satellite galaxies is a proxy for the dark matter halo mass corresponding to the time of infall. 

To apply and interpret Equation \ref{eq:sigma_dm} appropriately, galaxies must be identified as centrals or satellites. Such identifications can be made using group catalogs \citep[e.g.,][]{Yang2007, Knobel2012, Tempel2014}. After this identification, the dark matter halo mass for quiescent galaxies can be calculated using Equation \ref{eq:sigma_dm}. 

The scatter in the relation between stellar velocity dispersion and dark matter halo mass is $\sim0.2$ dex (see Figure \ref{fig:v3}B). This scatter is comparable with the scatter in the relation between stellar mass and halo mass determined from abundance matching \citep[and references therein]{Behroozi2013a}. However, an important difference is that the relation we derive in Equation \ref{eq:sigma_dm} is based solely on simulations and does not require matching to observations. {This result can be tested observationally. For example, Utsumi et al. (2018, in preparation) perform a joint spectroscopic and weak lensing observational analysis of the relation between stellar velocity dispersion and dark matter halo velocity dispersion.}

\section{Discussion}

{Observational results indicate that the central stellar velocity dispersion of a galaxy is a robust proxy of dark matter halo velocity dispersion \citep{Schechter2015, Zahid2016c}. We examine the Illustris-1 hydrodynamical simulation to explore this connection between stellar velocity dispersion and dark matter halo properties.}

Dark matter halos are in virial equilibrium resulting in a tight correlation between the mass and velocity dispersion of the halo. Line-of-sight aperture stellar velocity dispersion---calculated in analogy to observations---is also tightly correlated with the dark matter halo velocity dispersion. Thus, stellar velocity dispersion scales with dark matter halo mass. This correspondence between the central stellar velocity dispersion and dark matter halo velocity dispersion was proposed by \citet{Schechter2015} and \citet{Zahid2016c}. 

{The correspondence between the stellar velocity dispersion and the dark matter halo velocity dispersion may not be surprising. Observational studies suggest that the total mass density profile of quiescent galaxies is very nearly isothermal, i.e., the mass density profile is $\propto r^{-2}$ where $r$ is the radial coordinate \citep[e.g.,][]{Gavazzi2007, Auger2010, Barnabe2011, Bolton2012b, Cappellari2015, Serra2016}. Massive early-type galaxies in Illustris are broadly consistent with these observational results \citep{Xu2017}. For an isothermal sphere the gravitational potential energy and velocity dispersion are independent of radius. In other words, if a galaxy mass distribution is very nearly isothermal, the stellar and dark matter motions trace the same gravitational potential irrespective of their radial distribution. Thus, we might expect that the stellar velocity dispersion of a quiescent galaxy is a robust proxy for the dark matter halo velocity dispersion \citep[see also][]{Schechter2015}.}

{The relation between stellar velocity dispersion and dark matter halo mass differs for central and satellite galaxies. \citet{Vale2006} and \citet{Conroy2006} discuss the need to account for subhalo mass loss when examining correlations between galaxies and their halos. Stellar velocity dispersion, unlike the dark matter halo, is robust to tidal effects. Thus, stellar velocity dispersion of satellite galaxies traces the dark matter halo mass at the time of infall. To strictly apply and interpret the relation between stellar velocity dispersion and dark matter halo mass we must distinguish between central and satellite galaxies. For nearby galaxies, central and satellite galaxies can be identified using group catalogs \citep{Yang2007, Knobel2012, Tempel2014}, though the identification is subject to significant uncertainty. At higher redshifts, the distinction is observationally challenging.} 

{Here we discuss potential links between the simulations and observations. We evaluate the impact of treating all objects as central galaxies in Section 4.1 and outline ways to select halos in N-body simulations which are consistent with the observations in Section 4.2. We outline a potential refinement to abundance matching in Section 4.3 and discuss how the simulation results may be observationally tested with galaxy-galaxy weak gravitational lensing in Section 4.4.}

\subsection{{Application to Redshift Surveys of Massive Galaxies}}

{Current spectroscopic surveys which measure velocity dispersions of galaxies outside the local universe, e.g., the Baryon Oscillation Spectroscopic Survey \citep[BOSS;][]{Eisenstein2011, Thomas2013, Maraston2013}, the Smithsonian Hectospec Lensing Survey \citep[SHELS;][]{Geller2014, Zahid2016c} and the Large Early Galaxy Census \citep[LEGA-C;][]{vanderwel2016}, target massive galaxies. Identifying central and satellite galaxies in these surveys is subtle. } 

{We assess the potential systematic error if central and satellite galaxies are not identified and all galaxies are treated as central galaxies. The source of the error is that the dark matter halo mass calculated from stellar velocity dispersion is systematically overestimated for satellite galaxies. Figure \ref{fig:sat}B shows the dark matter halo mass loss due to stripping of satellites; the median fractional mass loss for the sample is 0.24 dex but the effect can be significantly larger for some dark matter halos.}

{Here we discuss the importance of distinguishing between central and satellite galaxies when applying the simulation results to current and future observations of quiescent galaxies. In Section 4.1.1 we investigate the selection of central quiescent galaxies in redshift surveys and in Section 4.1.2 we explore the application of our results to statistical analyses of redshift surveys.}

\subsubsection{Selection of Central Galaxies in Redshift Surveys}

\begin{figure*}
\begin{center}
\includegraphics[width =  1.82\columnwidth]{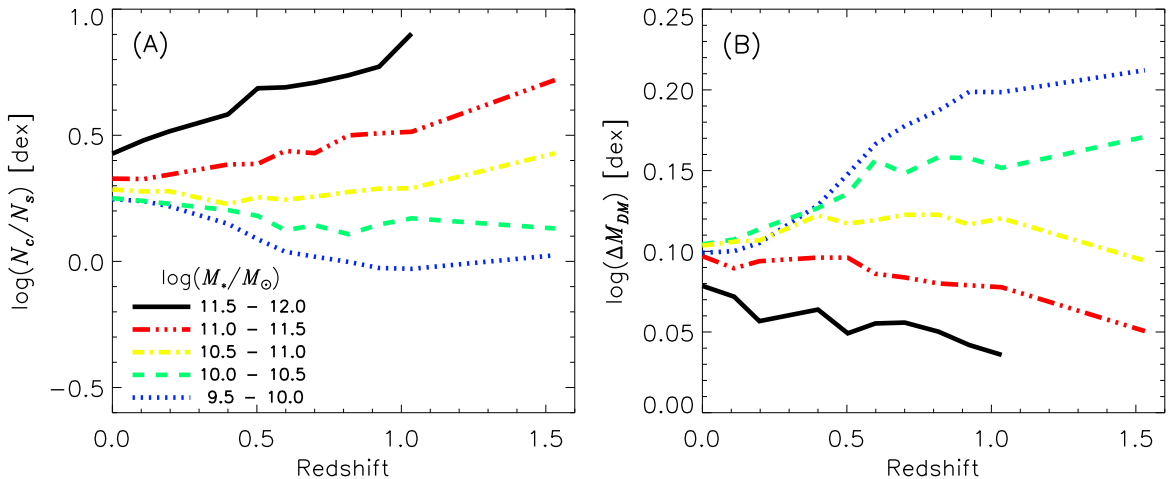}
\end{center}
\caption{(A) Ratio of the number of central galaxies to satellite galaxies as a function of stellar mass and redshift. We examine the quiescent galaxy population at several redshift slices with $\Delta z \sim 0.1$ and in stellar mass bins of 0.5 dex. The inlay denotes the stellar mass bin corresponding to the different color curves and line styles. At $z>1$ there is a dearth of galaxies in the highest mass bin. (B) Fractional error in dark matter halo mass as a function of stellar mass and redshift. We assume central and satellite galaxies are not identified and the sample is analyzed as a statistical ensemble. Details of the calculation are given in the text.}
\label{fig:zmass}
\end{figure*}

{We assess the selection of central and satellite galaxies in redshift surveys by calculating the ratio of the number of central to satellite galaxies in the simulation as a function of stellar mass and redshift. We carry out the calculation at several snapshots with $\Delta z \sim 0.1$ and in stellar mass bins of 0.5 dex.} 

{Figure \ref{fig:zmass}A shows the ratio of the number of central galaxies, $N_c$, to satellite galaxies, $N_s$, as a function of redshift and stellar mass. The central galaxy fraction increases as a function of stellar mass and for massive galaxies as a function of redshift. For galaxies with $M_\ast > 10^{10.5} M_\odot$, $\gtrsim 2/3$ of the population are central galaxies. At $z\sim1$ almost all ($\sim 90\%$) massive quiescent galaxies are central galaxies.}

{The dominance of central galaxies in the massive quiescent galaxy population has important implications for analyzing large samples of galaxies in magnitude limited surveys.}

\subsubsection{Application to Magnitude Limited Surveys}

{A standard approach to analyzing observations is to stack data or calculate statistical properties of a sample of galaxies. For example, \citet{Zahid2016c} examine the average relation between stellar mass and stellar velocity dispersion for massive quiescent galaxies at $z<0.7$ and Utsumi et al. (2018, in preparation) measure the average relation between stellar velocity dispersion and dark matter halo velocity dispersion by combining spectroscopy and weak lensing. The properties analyzed in these studies---stellar mass, stellar velocity dispersion and dark matter halo velocity dispersion---are insensitive to stripping and thus a distinction between central and satellite galaxies is not necessary. However, calculating dark matter halo mass from the average stellar velocity dispersion is subject to systematic uncertainty when central and satellite galaxies are not distinguished.}

{We estimate the systematic error in dark matter halo mass derived using the stellar velocity dispersion measured from stacked or sample averaged observations. As before, we calculate the systematic error as a function of redshift and stellar mass treating all objects as central galaxies. The systematic error estimate for each stellar mass bin and snapshot is
\begin{equation}
\Delta M_{DM} = \frac{ (N_c + N_s) ~ \overline{M_{DM, c}} }{  \sum_{i=1}^{N_c + N_s} M_{DM,i}  }.
\label{eq:dm}
\end{equation}
Here, $\overline{M_{DM, c}}$ is the mean dark matter halo mass of central galaxies for a single snapshot and stellar mass bin. The numerator of Equation \ref{eq:dm} is the total dark matter mass calculated assuming all galaxies in the bin have the mean dark matter halo mass of central galaxies in that bin. The denominator is the true total dark matter halo mass summed over all galaxies in the bin---including satellite galaxies which may be stripped. Thus, $\Delta M_{DM}$ is the systematic overestimate in the dark matter halo mass one would calculate from a sample average velocity dispersion assuming all galaxies are central galaxies.}

{Figure \ref{fig:zmass}B quantifies the systematic overestimate in dark matter halo mass when galaxies are analyzed as a statistical ensemble. For galaxies with $M_\ast > 10^{10.5} M_\odot$, the systematic uncertainty is $\lesssim 0.1$ dex. At $z\gtrsim1$ the effect is $\lesssim 0.05$ dex for massive ($M_\ast > 10^{11} M_\odot$) quiescent galaxies. These small systematic errors reflect the dominance of central galaxies in the massive quiescent galaxy population shown in Figure \ref{fig:zmass}A.}

{Current galaxy surveys like BOSS, SHELS and LEGA-C target massive galaxies at $z\lesssim1$. For these surveys, $\gtrsim 2/3$ of galaxies are central galaxies. If these survey data are analyzed statistically, the systematic error in deriving dark matter halo mass from stellar velocity dispersion is typically $\lesssim 0.1$ dex. Thus, we conclude that the impact of systematic error resulting from not distinguishing between central and satellite galaxies is small for massive quiescent galaxies. The small systematic uncertainties greatly simplify the application of the simulation results to currently available observations.}

\subsection{Selecting Quiescent Galaxies} 

We analyze quiescent galaxies because their stellar kinematics are typically dominated by random motions \citep[e.g.,][]{Brinchmann2004}; a necessity for interpreting central stellar velocity dispersion as a virial quantity. There are various approaches for selecting quiescent galaxies in the literature \citep{Moresco2013}. The $D_n4000$ index is a directly measured spectroscopic proxy of stellar population age \citep{Balogh1999, Kauffmann2003a}; it can be used to classify quiescent galaxies \citep[e.g.,][]{Zahid2016a, Zahid2016c, Zahid2017a, Sohn2017a, Sohn2017b}. \citet{Damjanov2018} demonstrate that the $D_n4000$ index is a robust classifier of quiescent galaxies and yields results very similar to color selection techniques. 

\citet{Zahid2017a} find that velocity dispersion, size, stellar mass and the $D_n4000$ index of galaxies are correlated.  At a fixed stellar mass, galaxies with large $D_n4000$ indices are smaller and have larger velocity dispersions. These trends indicate that galaxies with older stellar populations have higher velocity dispersions and are more compact ostensibly reflecting fundamental properties of their dark matter halos. 

Identifying dark matter halos hosting quiescent galaxies is important for connecting observations and theory. We select quiescent galaxies in the Illustris simulations by making a cut in specific star formation rate. However, for dark matter only simulations this type of selection is impossible. Stellar population age appears to be correlated with the age of the dark matter halo \citep{Hearin2013}. Thus, the dark matter halo age can be used to select quiescent galaxies in analogy to observational use of the $D_n4000$ index. Dark matter halo age is also correlated with other properties of the dark matter halo, e.g., halo concentration \citep{Wechsler2002, Napolitano2010, vandenBosch2014}. Analyses connecting galaxies to dark matter halos in pure N-body simulations may identify halos likely to host quiescent galaxies by using dark matter halo age or properties such as halo concentration which are correlated with dark matter halo age.

\subsection{Abundance Matching and Velocity Dispersion}

Abundance matching is a standard technique for linking galaxies to their dark matter halos in simulations. The simplest abundance matching model associates the most luminous observed galaxy in a survey with the most massive halo in a simulated volume of equal size, the second most luminous galaxy with the second most massive halo and so on \citep[e.g.,][]{Vale2004, Kravtsov2004}. \citet{Shankar2006} modify the standard abundance matching approach by using various galaxy observables. For example, they associate galaxies with dark matter halos assuming a monotonic relation between stellar velocity dispersion and dark matter halo mass \citep[see also][]{Chae2011}. \citet{Chae2012} develop this approach further making a bivariate statistical match of stellar velocity dispersion and stellar mass to dark matter halo mass. Several works have suggested matching galaxies to halos using two observable parameters and two halo parameters \citep{Hearin2013, Masaki2013, Kulier2015, Saito2016}.

\begin{figure}
\begin{center}
\includegraphics[width =  \columnwidth]{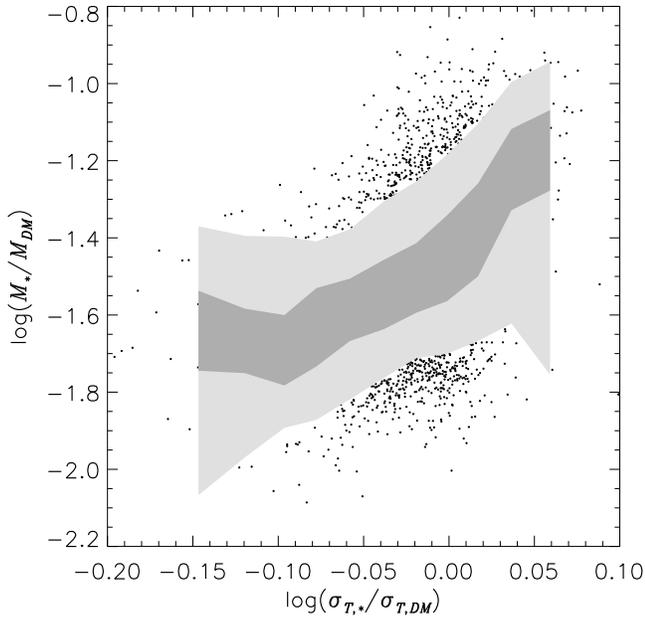}
\end{center}
\caption{Stellar-to-halo mass ratio as a function of stellar-to-halo velocity dispersion ratio for central galaxies. Scatter in the two relations is significantly correlated; the Spearman's rank correlation coefficient is $\rho = 0.42$.}
\label{fig:res}
\end{figure}

We examine whether inclusion of velocity dispersion is useful in reducing the $\sim 0.2$ dex scatter in standard abundance matching relations \citep{Yang2009, Behroozi2010, More2011}. Figure \ref{fig:res} shows that the stellar-to-halo mass ratio is correlated with the stellar-to-halo velocity dispersion ratio. Figure \ref{fig:res} suggests that for quiescent galaxies, stellar velocity dispersion is an important observable which can reduce the scatter in abundance matching relations.

Unlike stellar mass, stellar velocity dispersion is a directly measured quantity related to the gravitational potential; systematic uncertainties in stellar velocity dispersion of quiescent galaxies are $<0.03$ dex \citep{Fabricant2013, Zahid2016a, Zahid2016c} as compared to the $\sim0.3$ dex uncertainties in stellar mass estimates \citep{Conroy2009a, Behroozi2010}. Our results suggest that stellar velocity dispersion is the primary observable for connecting galaxies and dark matter halos; stellar mass may be used as a second parameter to help reduce the scatter. 

\subsection{Observational Tests with Weak Gravitational Lensing}

Galaxy-galaxy weak gravitational lensing is an observational approach for testing the Illustris results \citep[for review of the technique see e.g.,][]{Hoekstra2008}. Galaxy-galaxy weak gravitational lensing can be applied to a statistical sample of galaxies to probe their dark matter halos to large radii. Testing the Illustris results requires a dense and complete spectroscopic survey that include measurements of stellar velocity dispersion and high-quality, deep imaging for weak lensing analysis.

\citet{vanUitert2013} investigate the relation between stellar mass and stellar velocity dispersion of galaxies and their dark matter halo mass as traced by weak gravitational lensing. They show that at small separations stellar mass and velocity dispersion both account for the lensing signal equally well. However, the two tracers contain independent information regarding the dark matter distribution. The conclusions of \citeauthor{vanUitert2013} are consistent with Figure \ref{fig:res} showing the correlation between the stellar-to-halo mass ratio and the stellar-to-halo velocity dispersion ratio. 

Utsumi et al. (2018, in preparation) also examine the relation between stellar velocity dispersion and dark matter halo velocity dispersion using Hyper Suprime-Cam imaging of the SHELS F2 spectroscopic survey field \citep{Geller2014}. They find that the stellar velocity dispersion is directly proportional to the lensing velocity dispersion derived assuming a singular isothermal profile. This result is consistent with results presented in Figure \ref{fig:virial}B. 

The \citeauthor{vanUitert2013} and Utsumi et al. results demonstrate the potential of combining high-quality spectroscopy and imaging for probing the connection between galaxies and their dark matter halos. This observational approach complements theoretical results from the Illustris simulation.

\section{Conclusion}

We examine the Illustris-1 hydrodynamical cosmological simulations and show that for quiescent galaxies, stellar velocity dispersion is proportional to dark matter halo velocity dispersion. Thus, we conclude that stellar velocity dispersion is a robust proxy of the dark matter halo mass and can be used to link galaxies to their dark matter halos. A major advantage of using stellar velocity dispersion as a dark matter halo proxy is that it can be directly measured with small systematic uncertainties.  

The relation between stellar velocity dispersion and dark matter halo mass differs for central and satellite galaxies. Dark matter halos of satellite galaxies have likely been tidally stripped; the central stellar velocity dispersion is insensitive to this effect. For satellite galaxies stellar velocity dispersion is a proxy for the dark matter halo mass at the time of infall. 

{The simulation results provide a means to connect the stellar velocity dispersion---a measurable property of galaxies---to the theoretical dark matter halo. Surveys such as BOSS, SHELS and LEGA-C measure stellar velocity dispersion for massive quiescent galaxies outside the local universe. However, distinguishing between central and satellite galaxies in these surveys is subtle.} 

{We examine the impact of treating all galaxies as central galaxies. A large fraction ($\gtrsim 2/3$) of massive quiescent galaxies are central galaxies and the systematic uncertainty in applying our relation to a statistical ensemble of velocity dispersion measurements is $\lesssim 0.1$ dex. The small systematic uncertainties simplify application of the simulation results to observations.}

The stellar-to-dark matter halo velocity dispersion is tightly correlated with the stellar-to-dark matter halo mass. This result implies that scatter in the stellar-to-halo mass relation derived from abundance matching could be significantly reduced if stellar velocity dispersion is used as an additional parameter.

Stellar velocity dispersion is a powerful directly observed property for connecting galaxies and dark matter halos. Galaxy-galaxy weak gravitational lensing is a complementary observational technique to probe the dark matter distribution. Wide-field multi-object spectrographs (e.g., Subaru PFS, VLT MOONS) combined with large high-resolution imaging surveys (e.g., Euclid, LSST) will be transformative for cosmology. They will deliver dense spectroscopy and deep, high-resolution imaging over large areas of the sky. Combining spectroscopy with gravitational lensing will be a premier technique of the next generation of cosmology. Future measurements will provide an unprecedented probe of the matter distribution which will be critical for understanding the nature of dark matter.

\acknowledgements
 
We thank Volker Springel, Paul Schechter and Dylan Nelson for critically reading the manuscript and providing useful comments. We also thank Paul Torrey and Dylan Nelson for assistance accessing the data and Paul Torrey, Mark Vogelsberger, Lindsay Oldham, Ho Seong Hwang, Peter Behroozi and Andi Burkert for helpful discussion. We are grateful to the Illustris Collaboration for making their data publicly available. HJZ acknowledges the generous support of the Clay Fellowship. MJG and JS are supported by the Smithsonian Institution.

\bibliographystyle{aasjournal}
\bibliography{/Users/jabran/Documents/latex/metallicity}

 \end{document}